\documentclass[12pt]{iopart}

\usepackage{graphicx}
\begin{document}

\title[Femtoscopic scales in central A+A collisions]{Femtoscopic scales in central A+A collisions at RHIC and LHC energies in hydrokinetic model}
\author{Yu.A. Karpenko$^{1}$, Yu.M. Sinyukov$^{1}$}

\address{$^{1}$ Bogolyubov Institute for Theoretical Physics, 14-b Metrologishna Str. Kiev 143, 03680 Ukraine }

\begin{abstract}
A study of the energy behavior of the interferometry radii is carried out for the RHIC and LHC energies within
the hydrokinetic model (HKM). The hydrokinetic
predictions for the HBT radii at LHC energies are compared with the recent results of the ALICE Collaboration. The role of
non-equilibrium and non-hydrodynamic stage of the matter evolution in formation of the femtoscopy scales at the LHC energies is analyzed. For this aim we develop the hybrid hydrokinetic model.\footnote{The talk presented by Yuri Sinyukov  at "`Quark Matter - 2011"', May 23-28 2011, Annecy, France  }
 
\end{abstract}


{\it Introduction}--- The first LHC results of the femtoscopy analysis in Pb+Pb collisions at $\sqrt{s}=2.76$ TeV has been published recently by the ALICE Collaboration \cite{Alice}.  The principal question is whether an understanding of the space-time matter evolution in Au+Au collisions at RHIC can be extrapolated to the LHC energies. Here we  present the quantitative predictions given for LHC  within hydrokinetic model (HKM)  earlier \cite{sin3}, compare them with the recent ALICE LHC results and make the corresponding inference based on the first results of new hybrid hydrokinetic model (hHKM).

{\it Hydrokinetic approach to A+A collisions}--- The detailed description of the hydrokinetic model (HKM) is done in Refs. \cite{PRL,PRC}.
It incorporates hydrodynamical expansion of the systems formed in
\textit{A}+\textit{A} collisions and their dynamical continuous decoupling
described by the escape probabilities. The HKM also is the correct basis  to switch over  a hydrodynamic evolution of continuous medium to an evolution of particles within cascade model like UrQMD \cite{sin4}. The model which matches the hydrokinetic model and UrQMD we call hybrid HKM (hHKM). 

Our results are all related to the
central rapidity slice where we use the boost-invariant Bjorken-like
initial condition. We suppose the proper time of thermalization of
quark-gluon matter to be
$\tau_0=1$ fm/c.  The initial energy density in the transverse plane
is supposed to be Glauber-like \cite{Kolb} with zero impact parameter. The maximal initial energy density -
$\epsilon(r=0)=\epsilon_0$ is the fitting parameter. From the analysis
of particle multiplicities we choose it for the top RHIC energy to be
$\epsilon_0 = 15$ GeV/fm$^3$. For the LHC energy $\sqrt{s}=2.76$ TeV the corresponding value is
$\epsilon_0 = 40$ GeV/fm$^3$ that in hydrokinetic model  corresponds to multiplicity of charged particles $dN_{ch}/d\eta \approx 1500$.
The pre-thermal transverse flow which the system already has at  $\tau_0=1$ fm/c \cite{sin1} is supposed to be linear in
radius $r_T$: $y_T=\alpha r_T/R_T$ where $\alpha$ is the second fitting parameter and
$R_T=\sqrt{<r_T^2>}$. From the best fit of the pion transverse
spectra at the RHIC energy we take $\alpha=0.28$ in HKM and $\alpha=0.18$ for hHKM and the same value we keep for the LHC energy. Note that the fitting parameter $\alpha$
absorbs also a positive correction for underestimated
resulting transverse flow especially in HKM where we do not account for the
the viscosity effects \cite{Teaney} (in hHKM these effects are effectively included  at the hadronic stage, therefore $\alpha$ is essentially less than in HKM).  

Following to Ref. \cite{PRC} we use at
high temperatures the EoS \cite{Laine} adjusted to the QCD lattice
data. We suppose the chemical freeze-out for the hadron
gas at $T_{ch}=165$ MeV \cite{PBM1}. Below $T_{ch}$ a
composition of the hadron gas is changed only due to resonance
decays into expanding fluid. We include 359 hadron states made of u,
d, s quarks with masses up to 2.6 GeV. The EoS in this non
chemically equilibrated system depends on particle number
densities $n_i$ of all the 359 particle species $i$:
$p=p(\epsilon,\{n_i\})$.

In the case of hHKM we provide a switch to UrQMD at the hypersurface $\tau(r) = \tau_{sw}= const$ where $\tau_{sw}$ is defined from the condition: temperature $T(\tau=\tau_{sw},r_T=0)= 165$ MeV. Then the non-equilibrium distribution functions in hadron resonance  gas calculated in HKM are the input distribution functions for UrQMD at $\tau=\tau_{sw}$.
\begin{figure*}[h!]
\centering
 \includegraphics[scale=0.71]{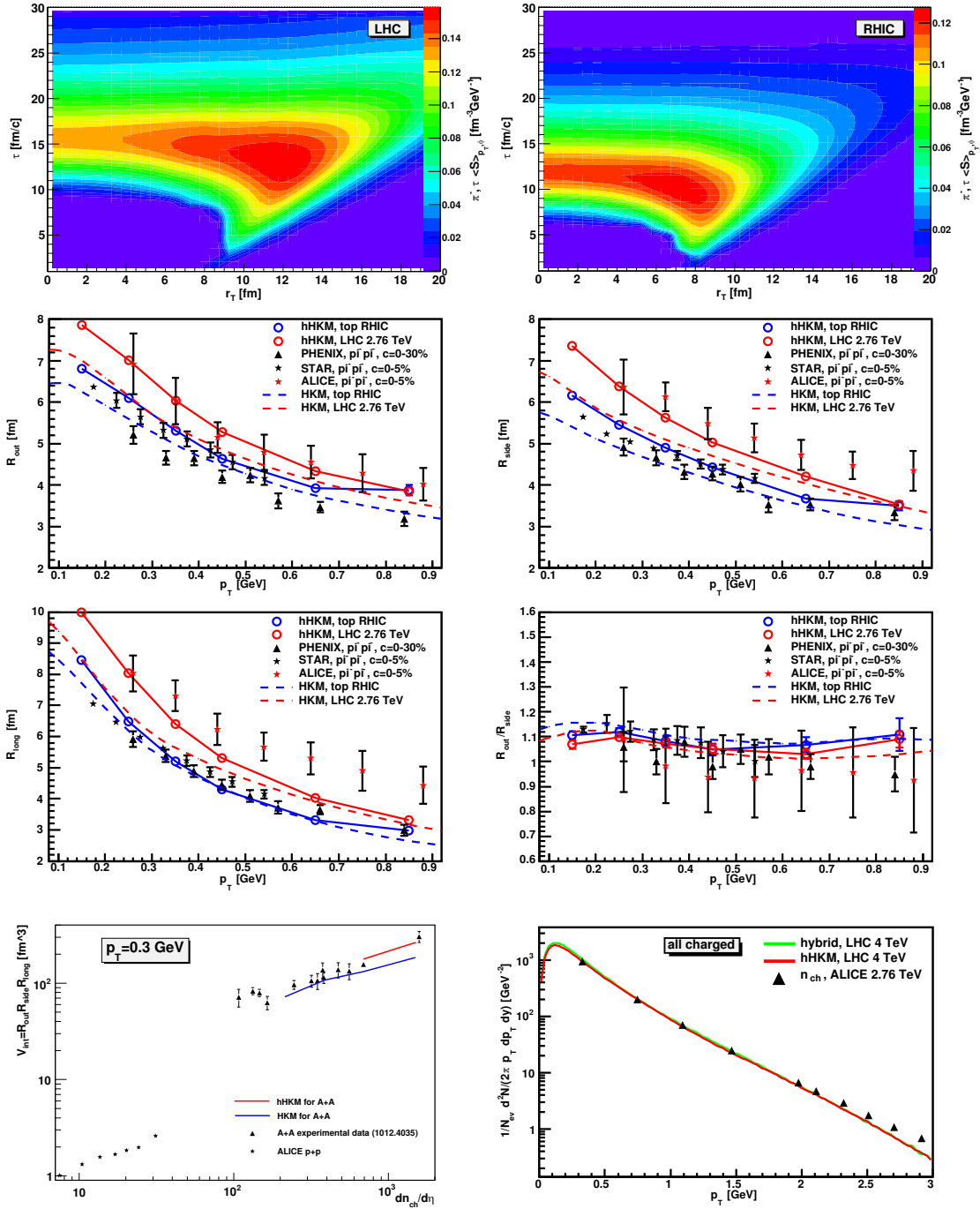}%

 \vspace{-0.19 in}
\caption{The $p_T$-integrated emission functions of negative pions
for the top RHIC and LHC energies in HKM(top); the interferometry
radii and $R_{out}/R_{side}$ ratio (middle); the transverse momentum
spectra  of all charged pions  at the LHC energy 2.76 TeV in hHKM model, and multiplicity dependence of the pion interferometry volume for central heavy ion collisions at  AGS, SPS, RHIC and LHC energies in comparison with the results in p+p collisions (bottom). The solid lines connect the points obtained within HKM and hHKM. The experimental data for A+A collisions are taken from
\cite{Alice, ceres, na49-spectra, na49-hbt,
star-spectra, star-hbt, phenix-spectra, phenix-hbt}. For p+p collisions at LHC the points for  $p_T=$ 0.3 GeV are interpolated from the results of Ref. \cite{Alice2}. }
\end{figure*}

{\it Results and conclusions}--- The pion emission functions per unit (central) rapidity, integrated
over azimuthal angular and transverse momenta, are presented in Fig.
1 for the top RHIC and LHC $\sqrt{s}=2.76$ TeV energies as a function of transverse
radius $r$ and proper time $\tau$. As one can see the duration of particle emission in the cental part of the fireball is
half of its lifetime. At a periphery a surface emission is significant, it lasts a total lifetime of the
fireballs. The hypersurface of the surface emission approaches the light cone when collision energy grows. The latter fact is correlated with a non trivial result on the energy behavior of the
$R_{out}/R_{side}$ ratio. It slowly drops when energy grows and
apparently is saturated at fairly high energies at the value close
to unity (Fig.1). In rough approximation  $R_{out}/R_{side}\approx 1+const/\epsilon_0$ \cite{sin3}.  This effect is caused by a strengthening  of
positive correlations between space and time positions of pions
emitted at the radial periphery of the system  as one can see in Fig.1, top (see  theoretical details in Ref. \cite{sin3}). This central prediction of  \cite{sin3} as for the energy dependence of the femtoscopy scales with energy growth is completely conformed by the ALICE results, see Fig. 1.

The transverse femtoscopy scales, predicted for LHC  are quite close (but slightly less) to the corresponding experimental data. As for the longitudinal HBT radius, $R_{long}$, it is underestimated  in HKM by around 20\%. As the result, HKM gives smaller interferometry volume than is observed at LHC, see Fig.1, bottom.  The reason could be that HKM describes a gradual decay of the system which evolves hydrodynamically until fairly large times. It is known \cite{AkkSin2} that at the isentropic and chemically frozen hydrodynamic evolution the interferometry volume increases quite moderate with initial energy density growth in collisions of the same/similar nucleus. The RHIC results  support such a theoretical view (see experimental and HKM results for SPS and RHIC, Fig1, bottom), while the ALICE Collaboration observes a significant increase of the interferometry volume at LHC.  An essential growth  of the interferometry volume in Pb+Pb collisons at the LHC energy is conditioned by a dominance of  very non-equilibrium and non-hydrodynamic stage of the matter evolution at LHC. We demonstrate it within the hHKM that containes such a stage, see Fig. 1.  Then the results obtained in \cite{AkkSin2} for isentropic and chemically frousen evolution are violated. Most significant effect is observed for the LHC energy. It is worthy noting that no one linear fit cannot be done to describe simultaneously  $V_{int}(dN/d\eta)$-dependence for both heavy (Pb, Au) ion and proton collisons, see for the latter the data discovered by the ALICE Collaboration  \cite{Alice2} (Fig. 1, bottom). This experimental observation supports the theoretical result  that the interferometry volume depends not only on multiplicity but also on intial size of colliding systems \cite{AkkSin2}. So, qualitatively, we see no new ``LHC HBT puzzle'' in the newest HBT results obtained at LHC in Pb+Pb and p+p collisions.

We conclude that the behavior of 
femtoscopy scales at LHC energy can be understood with the same hydrokinetic basis as was used for RHIC \cite{sin3,sin4} supplemented by hadronic cascade model at the latest stage of the evolution: HKM $\rightarrow$ hHKM. In this approach the following factors are important:  a presence of prethermal transverse flow, a crossover transition between quark-gluon and hadron matters, non-hydrodynamic behavior of the hadron gas at the latest stage, and correct matching between hydrodynamic and non-hydrodynamic stages. Then the particle spectra together with the femtoscopy data at RHIC and LHC energies can be well described.

{\it Acknowledgments}--- The researches were carried out within GDRE: Heavy ions at ultrarelativistic energies and is supported by  NAS of Ukraine (Agreement No F8-2011) and State Fund for Fundamental Researches of Ukraine (Project No. F33.2/001).

\end{document}